\documentstyle[epsfig]{aipproc}
\def\lsim{\mathrel{\raise.3ex\hbox{$<$\kern-.75em\lower1ex\hbox{$\sim$}}}}
\def\gsim{\mathrel{\raise.3ex\hbox{$>$\kern-.75em\lower1ex\hbox{$\sim$}}}}

    \def\fillboxx#1#2{\hbox to #1{\vbox to #2{\vfil}\hfil}    }

\def\tanb{\tan\beta}

\def\mt{m_t}

\def\mz{m_Z}
\def\mw{m_W}

\def\wp{W^+}
\def\wm{W^-}

\def\h{h}
\def\mh{m_{\h}}
\def\H{H}

\def\hpm{H^{\pm}}
\def\mhpm{m_{\hpm}}

\def\rts{\sqrt s}

\def\gam{\gamma}
\def\anti{\overline}
\def\epem{e^+e^-}

\def\hsm{h_{SM}}
\def\mhsm{m_{\hsm}}

\def\hl{h^0}
\def\mhl{m_{\hl}}

\def\ha{A^0}
\def\mha{m_{\ha}}

\def\hh{H^0}
\def\mhh{m_{\hh}}
\def\hp{H^+}
\def\hm{H^-}

\def\dchisq{\Delta\chi^2}
\def\dchisqmin{\dchisq_{\rm min}}

\def\abi{~{\rm ab}^{-1}}
\def\ab{~{\rm ab}}

\def\mev{~{\rm MeV}}
\def\gev{~{\rm GeV}}
\def\tev{~{\rm TeV}}

\def\MPL #1 #2 #3 {{\sl Mod.~Phys.~Lett.}~{\bf#1} (#3) #2}
\def\NPB #1 #2 #3 {{\sl Nucl.~Phys.}~{\bf #1} (#3) #2}
\def\PLB #1 #2 #3 {{\sl Phys.~Lett.}~{\bf #1} (#3) #2}
\def\PR #1 #2 #3 {{\sl Phys.~Rep.}~{\bf#1} (#3) #2}
\def\PRD #1 #2 #3 {{\sl Phys.~Rev.}~{\bf #1} (#3) #2}
\def\PRL #1 #2 #3 {{\sl Phys.~Rev.~Lett.}~{\bf#1} (#3) #2}
\def\RMP #1 #2 #3 {{\sl Rev.~Mod.~Phys.}~{\bf#1} (#3) #2}
\def\ZPC #1 #2 #3 {{\sl Z.~Phys.}~{\bf #1} (#3) #2}
\def\IJMP #1 #2 #3 {{\sl Int.~J.~Mod.~Phys.}~{\bf#1} (#3) #2}
\def\NIM #1 #2 #3 {{\sl Nucl.~Inst.~and~Meth.}~{\bf#1} {#3} #2}
\def\JHEP #1 #2 #3 {{\sl JHEP}~{\bf#1} (#3) #2}
\newcommand{\nc}{\newcommand}
\nc{\beq}{\begin{equation}}   \nc{\eeq}{\end{equation}}
\nc{\bea}{\begin{eqnarray}}   \nc{\eea}{\end{eqnarray}}
\nc{\baa}{\begin{array}}      \nc{\eaa}{\end{array}}
\nc{\bit}{\begin{itemize}}    \nc{\eit}{\end{itemize}}
\nc{\ben}{\begin{enumerate}}  \nc{\een}{\end{enumerate}}
\nc{\bce}{\begin{center}}     \nc{\ece}{\end{center}}
\def\beqa{\begin{eqnarray}}
\def\eeqa{\end{eqnarray}}

\begin{document}
\title{Do precision electroweak constraints guarantee \boldmath$\epem$
collider discovery
of at least one Higgs boson of a type-II two-Higgs-doublet model?
\thanks{Supported
by the U.S. Department of Energy and U.C. Davis.
Much of the work summarized here was performed in collaboration
with P. Chankowski, T. Farris, B. Grzadkowski, J. Kalinowski, and 
M. Krawczyk.}
\thanks{To be published in the Proceedings of LCWS2000, 
Fermilab Linear Collider Workshop, October 24-28, 2000.}}

\author{John F. Gunion}
\address{Davis Institute for High Energy Physics\\
Department of Physics, University of California, Davis, CA 95616}
%National Center for Atmospheric Research
%\thanks{The National
%Center for Atmospheric Research is sponsored by the National
%Science Foundation.}\\
%Boulder Colorado 80307\\
%$^{\dagger}$National Standards Institute, Boulder, Colorado 11543}

%\lefthead{LEFT head}
%\righthead{RIGHT head}
\maketitle

\vskip -.2in
\begin{abstract}
The manner in which the parameters of a two-Higgs-doublet model can
be chosen so that no Higgs boson is discovered at a $\rts\leq 800\gev$
$\epem$ collider, while maintaining consistency with current
precision electroweak measurements, is described. The importance
of a Giga-$Z$ factory and higher collider 
energies for such a scenario is emphasized. 
\end{abstract}

Models abound in which the 2HDM extension of the SM (without
supersymmetry) is the effective theory, 
correct up to some new physics scale, $\Lambda$
(set by, for example, the size of extra dimensions or the scale
at which new interactions creating the Higgs bosons as
bound states of fermions become strong). 
In general, the 2HDM can be CP-violating. However, for simplicity we focus on
the CP-conserving 2HDM of type II, with eigenstates
$\hl$, $\hh$, $\ha$ and $\hpm$.
Unlike the MSSM 2HDM, the
quartic Higgs potential couplings, $\lambda_i$, are not a priori
determined by gauge couplings. Further, the $\lambda_i$ need
only remain perturbative up to $\Lambda$,
which could easily be $\sim 1-10\tev$.
2HDM Higgs boson masses up to $\sim 1$ TeV are
consistent with $\alpha_i\equiv\lambda_i^2/(4\pi)$ below ${\cal O}(1)$
at such $\Lambda$. It is only if 
the $\alpha_i$ are required to remain $\lsim 1$ after evolving
to $\Lambda\sim 10^{19}\gev$ that the 2HDM predicts
$90\gev\lsim \mhl\lsim 175\gev$ and near maximal $g_{ZZ\hl}$ coupling
\cite{okada}. More generally, the decoupling limit 
\cite{decoupling},
defined by $\alpha_i\sim{\cal O}(1)$, 
$\mhh,\mha,\mhpm\sim M$  large (in which limit the $\hl$ is SM like),
need not be nature's choice.
Here, we focus on non-decoupling scenarios and summarize
the extent to which a $\sqrt s\sim 500-800\gev$
$\epem$ collider with high luminosity (we adopt a benchmark of $L=1\abi$)
is guaranteed to discover at least one 
Higgs boson. We also discuss the extent to which
the complementary Giga-$Z$ factory
would find clear direct or indirect evidence of a 2HDM sector, particularly
in cases where no Higgs boson would be discovered in $\rts\sim 500-800\gev$
running. 

The greatest strength of an $\epem$ collider
is that a light Higgs boson with any significant
$ZZ$ coupling will be discovered. 
But, for moderate $\tanb$,
the one-loop induced $ZZ\h$ coupling \cite{ghsmall} of
a light $\h$ with no tree-level $ZZ$ coupling (i.e. $\h=\ha$ or 
$\h=\hl$ with $\sin(\beta-\alpha)=0$,
the latter implying a special symmetry for the Higgs potential)
is sufficiently small that $Z\to Z^*\h$ 
would not have been detected at LEP and would not be detected
at a Giga-$Z$ factory. The alternative $Z\to\h\gam$ decays 
would also not be detected if $0.3<\tanb<15$ \cite{kzm}.
At large $\rts$, the off-shell behavior of the one-loop
induced  $Z^*Z\h$ coupling suppresses the cross section  $\propto\mt^2/s$ 
and $\sigma^{\rm 1-loop}(Z\h)$ falls below the level needed for 20 events
at $L=1\abi$ unless $\tanb<1$. 
For example, for $\mha^2\ll\mt^2,s$, the dominant $t$-loop diagram gives 
$\sigma^{\rm 1-loop}(Z\ha)\times \tan^2\beta\sim 3.6
\ab$ ($\sim 1.6\ab$) 
at $\rts=500\gev$ ($800\gev$).\footnote{Although $WW$ fusion does not
have $1/s$ suppression, the 1-loop induced cross section is
also too small at these energies.} 

Thus,  alternative production mechanisms for a light $\h$
with $g_{ZZ\h}=0$ should be considered.  If 
$\ha$ and $\hl$ are both light, then 
$\ha\hl$ production ($\propto \cos^2(\beta-\alpha)$)
and $Z\hl$ production ($\propto \sin^2(\beta-\alpha)$) cannot be 
simultaneously suppressed. If both the CP-even Higgs bosons, $\hl$ and $\hh$,
 are light, then $Z\hh$ production ($\propto \cos^2(\beta-\alpha)$)
will be detected if $Z\hl$ production is not. Current LEP limits
on the masses of two light Higgs bosons are 
significant \cite{twolight}, and would
be correspondingly increased at higher $\rts$.
If there is only one light Higgs boson, $\h$, and it has zero
$ZZ\h$ coupling, the four important  
production mechanisms are: $\epem\to \h\h Z$, $b\anti b \h$,
$t\anti t\h$ and $\gam\gam\to \h$.

\begin{figure}[h]
\begin{center}
\includegraphics[height=12cm]{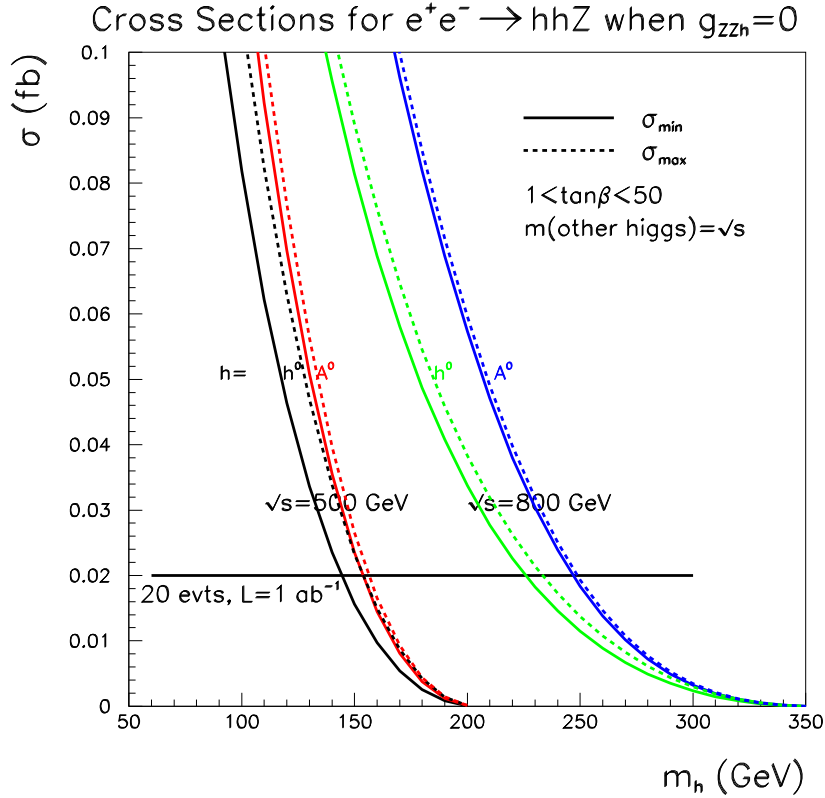}
\vspace*{-.2in}
\caption{
For $\protect\rts=500\gev$  and $\protect\rts=800\gev$ and for $\h=\hl$
and $\h=\ha$,
we plot as a function of $\mh$ the maximum and minimum values of
$\sigma(\epem\to \h\h Z)$  found after scanning $1<\tanb<50$
taking all other higgs masses equal to $\rts$. For $\h=\hl$, we
require $\sin(\beta-\alpha)=0$ during the scan. The 20 event level
for $L=1\abi$ is indicated.}
\label{hhz}
\end{center}
\end{figure}

The $\h\h Z$ and $\wp\wm\to \h\h$ processes \cite{gang,habernir,djouadietal}
are guaranteed to have contributions from the purely
gauge $\h\h ZZ$ and $\h\h \wp\wm$ couplings, respectively.  The additional
diagrams involving Higgs exchanges tend to suppress the cross section
relative to the purely gauge-coupling result.
Figure~\ref{hhz} gives the maximum and minimum values
of $\sigma(\h\h Z)$ obtained after scanning $1<\tanb<50$
for both $\h=\ha$ and $\h=\hl$
(with $\sin(\beta-\alpha)=0$ in the latter case) at $\rts=500$ and $800\gev$.
Other Higgs masses are taken equal to $\rts$. Assuming (optimistically)
that 20 events
will be adequate for discovery with $L=1\abi$, $\mhl<145\gev$ ($<225\gev$)
and $\mha<155\gev$ ($<250\gev$) will be probed at $\rts=500\gev$ ($800\gev$).

\begin{figure}[h]
\begin{center}
\includegraphics[height=11cm]{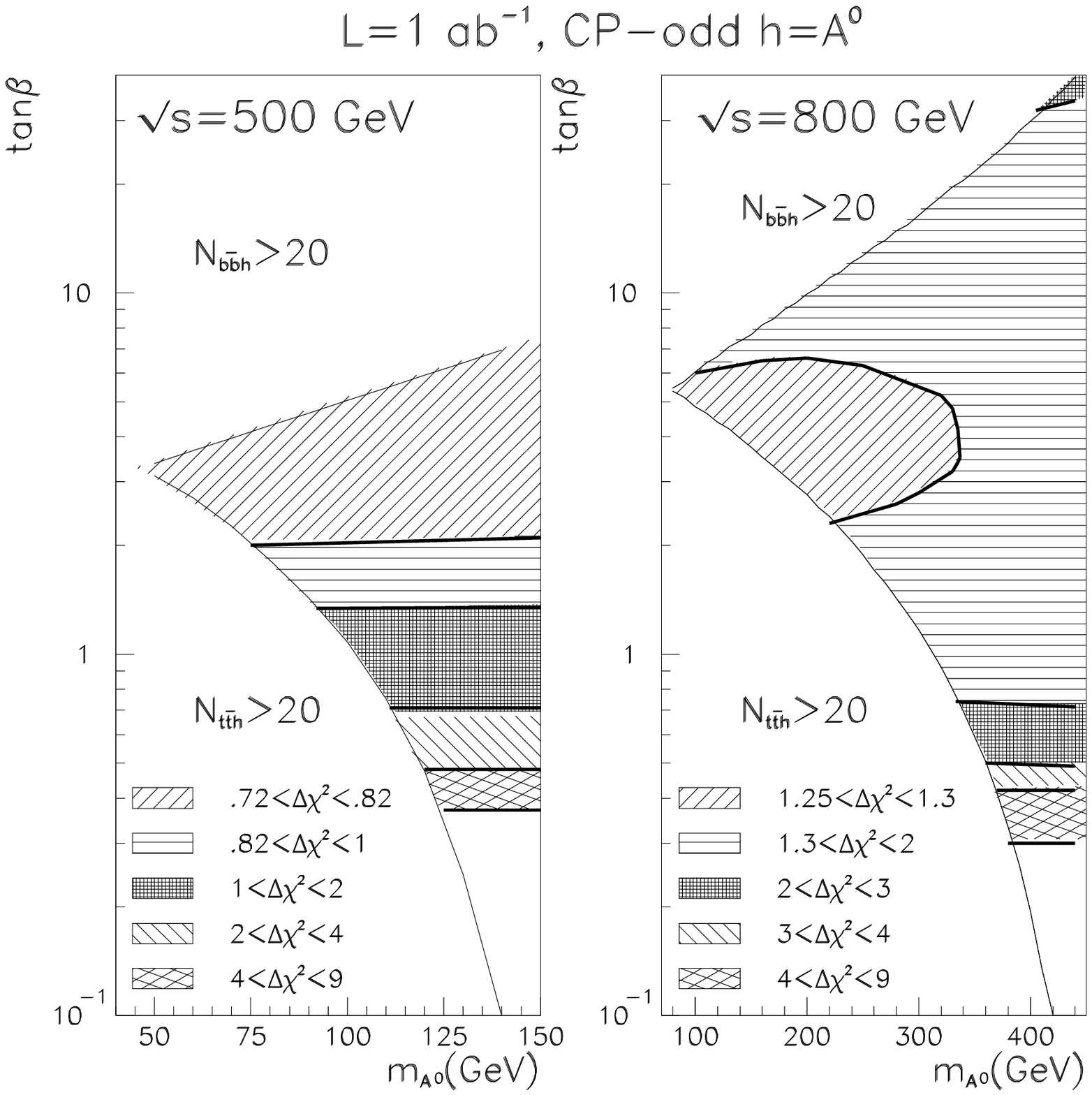}
\vspace*{-.2in}
\caption{
For $\protect\rts=500\gev$  and $\protect\rts=800\gev$,
the solid lines show
as a function of $\mha$ the maximum and minimum $\tanb$
values between which $t\anti t \ha$, $b\anti b \ha$ 
final states will both have fewer than 20 events assuming
$L=1\abi$. The different regions indicate
the best $\dchisq$ values (relative to the best SM $\chi^2$) 
obtained for fits to precision electroweak
data after scanning: a) over the masses of the remaining Higgs bosons 
subject to the constraint they are 
too heavy to be directly produced; and b) over the mixing angle
in the CP-even sector. Results are shown only for
$\mh<\protect\rts-2\mt$, but extrapolate to higher $\mh$ in obvious fashion.}
\label{lctdr_regions}
\end{center}
\end{figure}

When $g_{ZZ\h}=0$, the Yukawa processes cannot both be coupling
suppressed. Sum rules guarantee \cite{ggk} that 
$[S^t_{\h}]^2+[P^t_{\h}]^2=\cot^2\beta$ and 
$[S^b_{\h}]^2+[P^b_{\h}]^2=\tan^2\beta$, where the fermionic Higgs coupling
is given by ${gm_f\over 2m_W}\anti f(S^f_{\h}+i\gamma_5 P^f_{\h})f\h$;
i.e. $S^f_{\h}$ and $P^f_{\h}$ are defined relative to canonical SM strength.
Still, there are large regions of $[\mh,\tanb]$ parameter
space for which the Yukawa process rates will be inadequate
for detection. These are shown in the case of $\h=\ha$ 
in Fig.~\ref{lctdr_regions} for $\mh+2\mt<\rts$. For $\tanb$ values
inside the wedge-shaped regions,
$N_{t\anti t\ha}<20$ {\it and} $N_{b\anti b\ha}<20$ for $L=1\abi$.\footnote{The 
wedges of $[\mha,\tanb]$ parameter space for which $b\anti b\ha$
and $t\anti t \ha$ will be undetectable
are essentially the same as found for a general CP-violating 2HDM $\h$
in \cite{ggk}.} At least we see that
if $\mh$ is small (e.g., for $\h=\ha$, 
$\mha \leq 50\gev$ at $\rts=500\gev$ or $\mha\leq 140\gev$
at $\rts=800\gev$)  the $\h$ will be detected, assuming 20
$b\anti b\h$ or $t\anti t\h$ events are sufficient. In contrast,
observation of $Z\to b\anti b\h$ at a Giga-$Z$ factory is not possible,
even for  $\mh<10\gev$, if $\tanb\lsim 7$ \cite{kzm}. 
Of course, the `no-discovery' wedges of Fig.~\ref{lctdr_regions} expand
considerably for $\mh>\rts-2\mt$;  
the $\tanb$ value below which  $b\anti b\ha$ production cannot be seen
continues to increase and there is no lower $\tanb$ bound to the wedges.
Similar results are obtained for 
$\h=\hl$ when the $ZZ\hl$ coupling is zero \cite{getal}. 
In these problematical regions of moderate $\tanb$
values and moderate $\mh$ masses where 
$\h$ production is kinematically possible but
$\h\h Z$, $t\anti t\h$ and $b\anti b\h$ production rates are too small
for observation, $\gam\gam\to \h$ production is also
unlikely to produce a detectable signal for expected luminosities.

\begin{figure}[h!]
\begin{center}
\includegraphics[height=14cm]{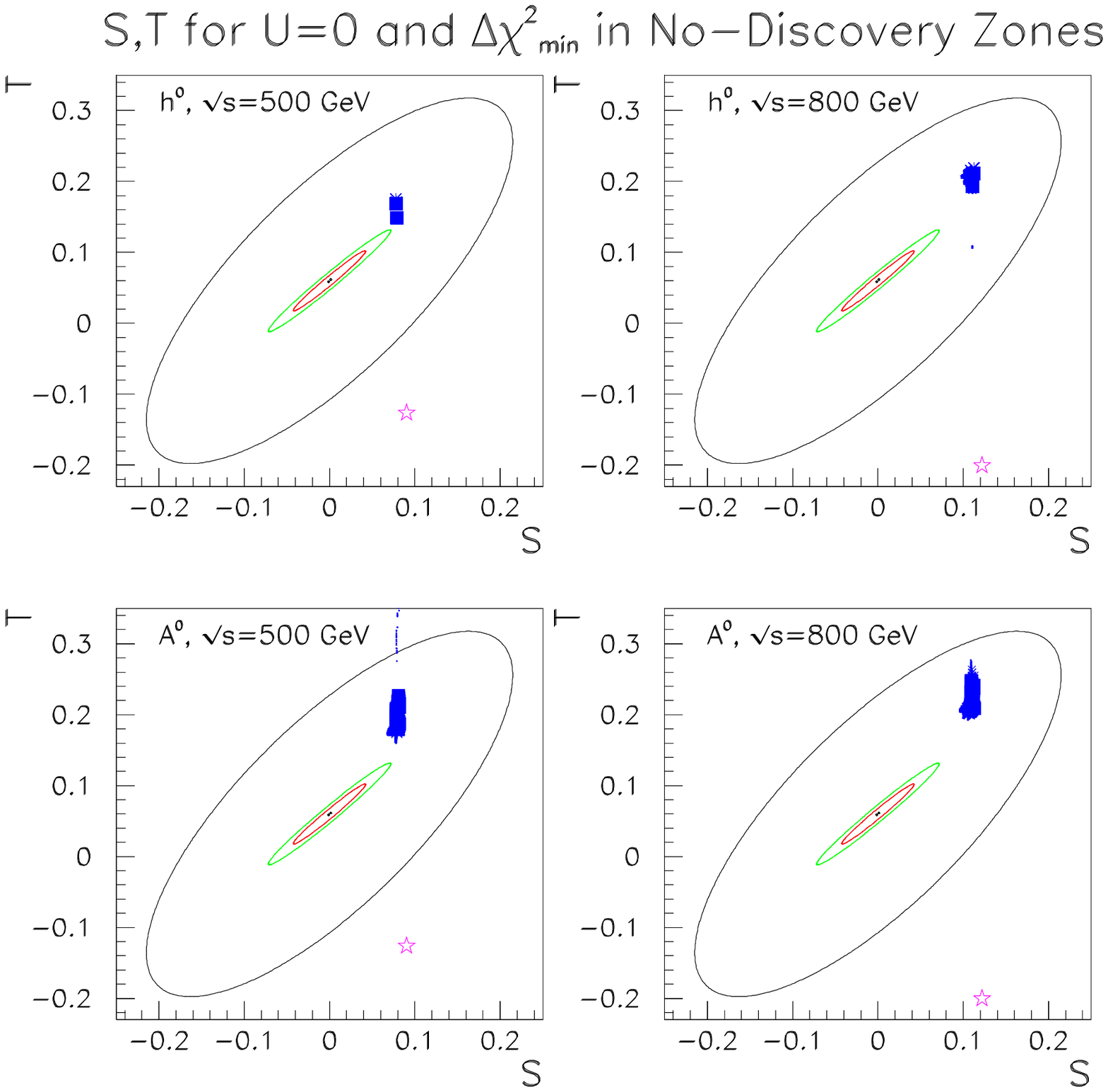}
\vspace*{-.2in}
\caption{The outer ellipse gives the current 90\% CL region
for $U=0$ and SM Higgs mass of 115 GeV.
$S,T$ predictions for the 2HDM $\dchisqmin$ models
with $[\mh,\tanb]$ values in the `no-discovery' wedges of 
Fig.~\protect{\ref{lctdr_regions}} are shown by the blobs of
overlapping points. The innermost (middle) ellipse gives the 90\% (99.9\%) CL
region for $\mhsm=115\gev$ obtained after
Giga-$Z$ precision measurements 
{\it and} a $\Delta m_W\lsim 6\mev$ threshold
scan measurement of $m_W$. The stars indicate the SM $S,T$ prediction if
$\mhsm=\rts-\mz$.
}
\vspace*{-.2in}
\label{dsdt}
\end{center}
\end{figure}

Surprisingly, in these scenarios, 
the parameters for the other (heavy) Higgs bosons can be chosen so
that the fit to precision electroweak observables is nearly as good
as that obtained with a light SM Higgs boson, despite the
fact that the CP-even Higgs boson with substantial
$WW,ZZ$ couplings is heavier than $\rts$ \cite{ggk,getal}.
This is illustrated
in Fig.~\ref{lctdr_regions} for the case of
$\h=\ha$. The $\Delta\chi^2$ values between the best 2HDM and SM
precision electroweak fits are seen to obey 
$\Delta\chi^2< 2$  in the $\rts=500\gev$ and
$\rts=800\gev$ $L=1\abi$  `no-discovery' wedges 
when $\tanb>0.7$. What is happening is best made clear using
the $S,T$ plane picture, Fig.~\ref{dsdt}.
The outer ellipse in Fig.~\ref{dsdt} defines the current
90\% CL region in the $S,T$ plane from precision measurements,  
assuming $U=0$ and a SM Higgs mass of $115\gev$ \cite{erler}. 
Focusing on the case of a light $\h=\ha$,
$\dchisqmin$ is achieved when: a) $\mhl\gsim \rts$; b) $g_{ZZ\hl}$ is
near maximum; and 
c) $\mhpm,\mhh>\mhl$ with $(\mhpm-\mhh)/\mhh$ small and positive.
If $\mhpm=\mhh$ exactly, then large $\mhl$ implies large 
positive (negative) contributions to $S$ ($T$) proportional to $\log(\mhl^2)$.
The predicted $S,T$ values would be 
essentially the same as for the SM with $\mhsm=\mhl$, and
would fall well outside the usual 90\% CL ellipse.
However, a small positive $\mhpm-\mhh$ gives a
new positive contribution to $T$ $\propto \mhpm^2-\mhh^2$ which can be
easily adjusted 
to move the central $S,T$ prediction back inside the ellipses.
The blobs of overlapping
points in Fig.~\ref{dsdt} are the $S,T$ predictions of the $\dchisqmin$ 
2HDM models for $[\mha,\tanb]$ values in the `no-discovery' 
zones of Fig.~\ref{lctdr_regions} and the analogous $[\mhl,\tanb]$ zones. 
Unless $\tanb<1$,
the predicted $S,T$ location is well within the current 90\% CL ellipse.

Significant improvement in
the precision of the electroweak measurements,
especially $\sin^2\theta_{\rm lep}^*$, would be achieved
at a Giga-$Z$ factory. If, in addition, a $\Delta m_W<6\mev$
measurement of $\mw$ is obtained by a $WW$ threshold scan (to verify
that $U$ is small to high precision), the resulting 90\% CL level ellipse
(plotted in Fig.~\ref{dsdt} assuming the SM with Higgs mass 115 GeV 
is correct) is greatly reduced,
as shown by the inner ellipse in Fig.~\ref{dsdt} \cite{erlermoenig}. 
If a $\dchisqmin$
`no-discovery' 2HDM scenario is nature's choice, the $S,T$ central value 
preferred by the precision measurements would 
be in the vicinity of the blobs. The SM, especially the version
of the SM with Higgs mass too heavy ($\mhsm>\rts-\mz$) for 
$Z\h$ production at the LC, predicts $S,T$ values
(the stars in Fig.~\ref{dsdt}) that would be clearly excluded.

In the $\dchisqmin$ `no-discovery' scenarios for $\h=\ha$ and $\h=\hl$, 
the mass of the CP-even Higgs $\H$ ($\H=\hl$ and $\H=\hh$, respectively)
with substantial $ZZ,WW$ coupling is 
never much beyond $1\tev$. 
The most probable future scenario would then be \cite{getal} discovery of
the $\H$ at the LHC (in the gold-plated $\H\to ZZ\to 4\ell$ mode)
and Giga-$Z$ measurements of $S,T$ that make clear the need
for a still heavier $\hpm$ and 2nd neutral Higgs boson
that are nearly, but not quite, degenerate in mass. There 
could still be a light $\h$. An LC with sufficient $\rts$ could
then completely reveal the Higgs states by observing not only $Z\H$
and/or $WW\to\H$ production
but also $\hl\ha$ production (regardless of which is light) and
possibly $\hp\hm$ production.


\begin{thebibliography}{99}

\bibitem{okada} 
%\cite{Kanemura:1999xf}:
%\bibitem{Kanemura:1999xf}
S.~Kanemura, T.~Kasai and Y.~Okada,
%``Mass bounds of the lightest CP-even Higgs boson in the  two-Higgs-doublet model,''
Phys.\ Lett.\  {\bf B471}, 182 (1999)
[hep-ph/9903289].
%%CITATION = HEP-PH 9903289;%%

%\cite{Haber:1994mt}:
\bibitem{decoupling}
%\bibitem{Haber:1994mt}
H.~E.~Haber,
%``Nonminimal Higgs sectors: The Decoupling limit and its phenomenological implications,''
hep-ph/9501320; J.F. Gunion and H.E. Haber, in preparation.
%%CITATION = HEP-PH 9501320;%%


\bibitem{ghsmall}
%\cite{Gunion:1992cw}:
%\bibitem{Gunion:1992cw} 
For the one-loop couplings see
J.~F.~Gunion, H.~E.~Haber and C.~Kao,
%``Searching for the CP odd Higgs boson of the minimal supersymmetric model at hadron supercolliders,''
Phys.\ Rev.\  {\bf D46}, 2907 (1992).
%%CITATION = PHRVA,D46,2907;%%


\bibitem{kzm}
%\cite{Krawczyk:1999kk}:
%\bibitem{Krawczyk:1999kk}
M.~Krawczyk, J.~Zochowski and P.~Mattig, hep-ph/0009201; see also
%``Process Z --> h(A) + gamma in the 2HDM and the experimental  constraints from LEP,''
Eur.\ Phys.\ J.\  {\bf C8}, 495 (1999)
[hep-ph/9811256].
%%CITATION = HEP-PH 9811256;%%

\bibitem{twolight}
%\cite{Gunion:1997aq}:
%\bibitem{Gunion:1997aq}
J.~F.~Gunion, B.~Grzadkowski, H.~E.~Haber and J.~Kalinowski,
%``LEP limits on CP-violating non-minimal Higgs sectors,''
Phys.\ Rev.\ Lett.\  {\bf 79}, 982 (1997)
[hep-ph/9704410]; for experimental analyses, see
P. Abreu {\it et al.}, DELPHI Collab., CERN-EP-2000-038 and
G. Abbiendi {\it et al.} OPAL Collab., CERN-EP-2000-092, [hep-ex/0007040].
%%CITATION = HEP-PH 9704410;%%

%\cite{Gunion:1988tf}:
\bibitem{gang}
%\bibitem{Gunion:1988tf}
J.~F.~Gunion {\it et al.},
%``Production Mechanisms For Nonminimal Higgs Bosons At An E+ E- Collider,''
Phys.\ Rev.\  {\bf D38}, 3444 (1988).
%%CITATION = PHRVA,D38,3444;%%

%\cite{Haber:1993jr}:
\bibitem{habernir}
%\bibitem{Haber:1993jr}
H.~E.~Haber and Y.~Nir,
%``The Decay Z $\to$ A0 A0 neutrino anti-neutrino and e+ e- $\to$ A0 A0 z in two Higgs doublet models,''
Phys.\ Lett.\  {\bf B306}, 327 (1993)
[hep-ph/9302228].
%%CITATION = HEP-PH 9302228;%%

%\cite{Djouadi:1999rc}:
\bibitem{djouadietal}
%\bibitem{Djouadi:1999rc}
A.~Djouadi, W.~Kilian, M.~Muhlleitner and P.~M.~Zerwas,
%``Production of neutral Higgs-boson pairs at LHC,''
Eur.\ Phys.\ J.\  {\bf C10}, 27 (1999)
[hep-ph/9903229].
%%CITATION = HEP-PH 9904287;%%



%\newcommand{\wwwspires}{http://www.slac.stanford.edu/spires/find/hep/www}
%\cite{Grzadkowski:1999ye}
\bibitem{ggk}
B.~Grzadkowski, J.~F.~Gunion and J.~Kalinowski,
%``Finding the CP-violating Higgs bosons at e+ e- colliders,''
Phys.\ Rev.\  {\bf D60}, 075011 (1999)
[hep-ph/9902308],
%%CITATION = HEP-PH 9902308;%%
%\href{\wwwspires?eprint=HEP-PH/9902308}{SPIRES}
%\newcommand{\wwwspires}{http://www.slac.stanford.edu/spires/find/hep/www}
%\cite{Grzadkowski:2000wj}
%\bibitem{Grzadkowski:2000wj}
%B.~Grzadkowski, J.~F.~Gunion and J.~Kalinowski,
%``Search strategies for non-standard Higgs bosons at future e+ e-  colliders,''
Phys.\ Lett.\  {\bf B480}, 287 (2000)
[hep-ph/0001093].
%%CITATION = HEP-PH 0001093;%%
%\href{\wwwspires?eprint=HEP-PH/0001093}{SPIRES}
See also
%\newcommand{\wwwspires}{http://www.slac.stanford.edu/spires/find/hep/www}
%\cite{Gunion:1997aq}
%\bibitem{Gunion:1997aq}
J.~F.~Gunion, B.~Grzadkowski, H.~E.~Haber and J.~Kalinowski,
%``LEP limits on CP-violating non-minimal Higgs sectors,''
Phys.\ Rev.\ Lett.\  {\bf 79}, 982 (1997)
[hep-ph/9704410].
%%CITATION = HEP-PH 9704410;%%
%\href{\wwwspires?eprint=HEP-PH/9704410}{SPIRES}

\bibitem{getal} 
%\cite{Chankowski:2000an}:
%\bibitem{Chankowski:2000an}
P.~Chankowski, T.~Farris, B.~Grzadkowski, J.~F.~Gunion, J.~Kalinowski and M.~Krawczyk,
%``Do precision electroweak constraints guarantee e+ e- collider  discovery of at least one Higgs boson of a two-Higgs-doublet model?,''
hep-ph/0009271;
%%CITATION = HEP-PH 0009271;%%
see also P.H.~Chankowski, M.~Krawczyk and J.~Zochowski, 
Eur.\ Phys. \. J. {\bf C11}, 661 (1999) [hep-ph/9608321].

\bibitem{erler} J. Erler and P. Langacker,  
www.\-physics.\-upenn.\-edu/\-\~\,erler/\-electroweak/\-results.html.

\bibitem{erlermoenig} We thank J. Erler and K. Moenig for communicating
their error estimates for this situation to us.


\end{thebibliography}
\end{document}